\documentstyle[12pt,epsf]{article}
\newcommand{\head}[1]{\vspace{\baselineskip} \noindent{\bf #1}\\}
\begin{document}
\begin{flushright}
RAL-TR-96-028 \\ 
hep-ph/9604413\\
\end{flushright}

\begin{center}
GAUGE UNIFICATION AT THE STRING SCALE AND FERMION MASSES

\vspace{2\baselineskip}
B.C.Allanach\footnote{Talk presented by B.C.Allanach at {\em XXXI
Rencontres de Moriond.}}, Rutherford Appleton Lab,
Didcot, Oxon., OX11 OQX, UK.\\
S.F.King, Dept.\ of Physics, University of Southampton, Southampton,
SO17 1BJ, UK.
\end{center}
\vspace{7cm}

In the context of the minimal supersymmetric standard model (MSSM), we
discuss the introduction of exotic matter below the string scale $M_X$
in order to achieve gauge unification at $M_X$ (a constraint of a
large class of string models). The possible types of exotic matter
that can realise this are investigated and its effect on the top quark
mass $m_t$ is presented. The implementation of a theory of fermion
masses which utilises the exotic matter is briefly discussed.

\newpage

\head{Gauge Unification}

\begin{figure}
\begin{center}
\leavevmode
\hbox{\epsfxsize=5in
\epsfysize=3in
\epsffile{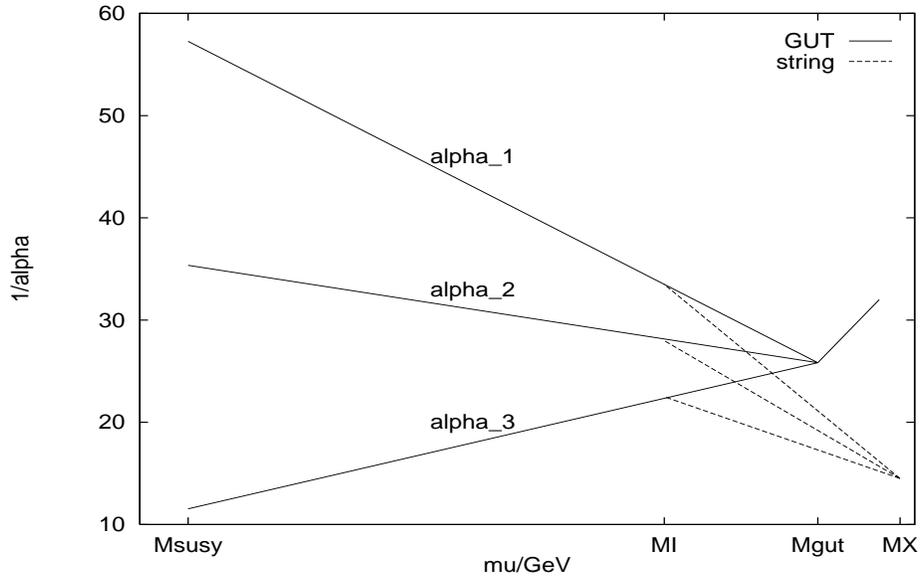}}
\end{center}
\caption{Evolution of the gauge couplings with energy scale
$\mu$ in the
context of a
supersymmetric GUT model and a string model with $k_1=5/3$ and extra
matter added
with mass $M_I$. The extra matter in this case is $2 \times(3,2)$ and
$3 \times (3,1)$ representations (and antiparticles).}
\label{fig:MSSM}
\end{figure}

It has been known for a long time that a constraint of simple
GUTs
such as SU(5), SO(10) is that the
gauge couplings should become equal when evolved up to
a high energy scale. This scale is the energy at
which the GUT breaks down into the Standard Model $M_{GUT}$ and this
``unification'' means that
\begin{equation}
\alpha_1(M_{GUT})=\alpha_2(M_{GUT})=\alpha_3(M_{GUT}),
\label{GUT}
\end{equation}
where $\alpha_{1,2,3}$ are the gauge couplings corresponding to the
U(1)$_Y$\footnote{$\alpha_1$ in Eq.~\protect\ref{GUT} assumes the GUT
normalisation 
of the hypercharge
$Y^{GUT}=\sqrt{3/5}Y^{SM}$.}, SU(2)$_L$ and SU(3) simple gauge groups
in the Standard
Model (SM)
respectively. 
Several authors have shown that Eq.~\ref{GUT} is not satisfied when
the effective theory at energy scales between $M_{GUT}$ and $M_Z$ is
the SM\@. The gauge couplings evolve because the gauge boson propagator
has a divergent contribution coming from loops of particles with
a mass less
than the relevant energy scale. This means that if
the particle content
of a theory 
is changed, so is the evolution of the gauge couplings. When
the SM is supersymmetrised to become the MSSM, the extra particles
(``superpartners'' and Higgs) alter the evolution in such a way that
the values of $\alpha_i(M_Z)$ that are extracted
from experiment and evolved to $M_{GUT}$ satisfy Eq.~\ref{GUT} to a
good accuracy$^{1)}$. This point is illustrated by
Fig.~\ref{fig:MSSM} in which the solid lines show the gauge couplings
meeting at $M_{GUT} \sim
10^{16}$~GeV. Simple GUTs only have one gauge coupling, which then
evolves at scales $\mu > M_{GUT}$ as shown\footnote{In the figure, the
running
is only shown from
$\mu=M_{SUSY}\sim1$~TeV but one should bear in mind that the input
values at $M_{SUSY}$ are the ones extracted from experiment and
evolved to $M_{SUSY}$ using the Standard Model renormalisation
group.}. 
While GUTs provide a simple and elegant scheme for helping to explain
the origin of the strong and electroweak forces, they do not include
any quantum description of gravity. The only known consistent theories
of quantum gravity to date are superstring theories, and we now turn to
these to examine how the above apparent success of gauge coupling
unification in GUTs translates into string models.

\head{String Gauge Boundary Conditions}

In string models, the constraints on the low energy\footnote{Low energy
in this paper refers to anything less than the string scale $M_X$.}
gauge couplings may be different to the ones in Eq.~\ref{GUT}. In
fact, the relevant relations are$^{2)}$ 
\begin{equation}
\frac{3}{5}k_1 \alpha_1(M_X) = k_2 \alpha_2(M_X) = k_3 \alpha_3(M_X)
\equiv
\alpha_{string} \label{strings}
\end{equation}
where $k_i$ are the Kac-Moody levels associated with the string model.
For our analysis, the important fact about $k_i$ is that they are
numerical constants that are set, once the string model has been
chosen. For non-abelian groups, the Kac-Moody levels must be natural
numbers whereas they are rational for abelian groups$^{3)}$. This
means that
$k_{2,3}\in \{ 1,2,3,\ldots \}$, $k_1 \in Z/Z$. The only semi-realistic
models
(i.e.\ with 3
families etc.) that have been constructed to date have $k_{2,3}=1$ and
so
we concentrate on this case.

At first sight it appears that one constraint is lost compared to the
GUT case because we have imposed no conditions upon $k_1$, the
normalisation of $Y$. However, string models with $k_{2,3}=1$ relate
the scale of the breakdown of string theory to the gauge couplings
through the relation
\begin{equation}
M_X = 5.3 \sqrt{4 \pi\alpha_{string}} \times 10^{17}
\label{scale}
\mbox{~GeV}.
\end{equation}
This would mean that low energy phenomenology is incompatible with the
prediction $\alpha_2(M_X)=\alpha_3(M_X)$, since if the gauge couplings
were evolved to $M_X$ in the MSSM, the solid lines in
Fig.~\ref{fig:MSSM} would still cross at $M_{GUT}$ and be different at
$M_X$.
There are, however several
reasons$^{4)}$ why the measured gauge couplings might (wrongly)
appear to be
in conflict
with Eq.~\ref{strings} and therefore with the class of string models
that we are
advocating. Several of these possible reasons have been
discussed by other
authors$^{4)}$, but
we turn to one particular possibility that unambiguously satisfies 
Eq.~\ref{GUT}.

\head{Intermediate Matter}

One can imagine a theory in which some matter additional to the
MSSM$^{3),4),5)}$
has a mass
$M_I$ where $M_Z < M_I < M_X$. Below $M_I$ we have the
MSSM, and above it the effect of the extra matter is felt upon the
gauge couplings. 
In string models which break to the MSSM and have $k_{2}=k_3=1$, the
only possible matter fields present in the low energy theory are
(3,1), (1,2) and (3,2) representations\footnote{Written in
(SU(3),SU(2)$_L$) space.}. Leaving hypercharge assignments aside,
these fields look like extra copies of right handed quarks $q_R$, left
handed lepton doublets $L_L$ and left handed quarks $Q_L$
respectively\footnote{It is to be understood that the supersymmetric
partners and conjugate (antiparticle) fields are to be automatically
included in the spectrum.}. We label the number of $q_R$, $L_L$ and
$Q_L$ fields $a,b,c$ respectively.
One can solve $\alpha_2(M_X)=\alpha_3(M_X)$ and Eq.~\ref{scale} to obtain
$M_I,M_X, \alpha_{string}$ 
for each
choice of possible intermediate matter. 
In Fig.~\ref{fig:MSSM}, the dotted lines show how the intermediate
matter can ``re-focus'' the gauge couplings to meet at the string scale
$M_X$. 
The bound $M_I < M_X$ implies that
\begin{equation}
a > b+c
\end{equation}
for the model to unify $\alpha_2(M_X)=\alpha_3(M_X)$.

\head{Top Quark Mass}

When one evolves the top quark Yukawa coupling $h_t$ from high to low
energy scales in the MSSM, one finds that the RGEs naturally ``focus''
the 
couplings into a narrow range at low energy centred around the infra
red stable fixed point (IRSFP)$^{6)}$.
This IRSFP of $h_t$
in the MSSM corresponds to $m_t / \sin \beta \approx 174-195$
GeV\footnote{$\tan \beta $ is the ratio of the two Higgs vacuum
expectation values (VEVs) of the MSSM.}. The ``quasi fixed point''
(QFP) is
defined when $h_t$ at some high energy scale (say
$M_X$) is large, and in this case the low energy value of the top mass
$m_t / \sin \beta \sim 210$ GeV is independent of what the actual
value of $h_t(M_X)$ is.
The IRSFP of the top quark Yukawa coupling is described analytically by
\begin{equation}
\left( \frac{h_t^2}{4 \pi \alpha_3} \right)^* \sim (16/3 + b_3)/6,
\end{equation}
where $b_3$ is the QCD beta function. In the region of energy scales
between $M_I$ and $M_X$, we have added coloured matter which
changes $b_3$ and could possibly change the low energy prediction of
the top quark mass.
Overall, the effect of the intermediate matter is to drive $h_t$
nearer to its QFP value that corresponds to $m_t / \sin \beta \approx
210$ GeV.

\head{Lighter Fermion Masses}

We now address the question of how the pattern and hierarchy of the
fermion masses and mixing angles may arise.
In our models, the only renormalisable fermion mass term is the one
in the 33 element
of the mass matrices, with all the others initially being zero. 
We now consider how the lighter fermions could acquire mass and mixing
angles.
One idea$^{7)}$ is to
extend the gauge symmetry of the
MSSM by a family dependent abelian U(1)$_X$. 
The theory contains SM singlets $\theta^1, \bar{\theta}^{-1}$ which
acquire VEVs, and these break the U(1)$_X$ symmetry. 
We introduce some heavy
Higgs doublets of mass $M$
with different $X$ charges and these help to generate
non-renormalisable operators which play the role of mass and mixing
terms. 
The
other entries in the matrix appear once $\theta$ acquires a VEV
$\langle \theta \rangle$.
The nonrenormalisable operators
produced in this way can
generate a predictive and explanatory scheme of masses and
mixings
angles$^{7),8)}$.

However, we have already been using the candidate heavy Higgs to help
unify 
$\alpha_i$ in string scenarios so we may be able to put these to work
to give a predictive theory of fermion masses. In this framework, the
U(1)$_X$ has mixed anomalies with SU(3), SU(2)$_L$, U(1)$_Y$ and to
cancel these 
using a (string-type) mechanism called GSW,
the Kac-Moody levels must be in the ratio 
\begin{equation}
k_1:k_2:k_3=5/3:1:1. \label{Ks}
\end{equation}
The GSW mechanism also requires $\langle \theta \rangle / M_X \ \sim
O(1/40)$ but for 
a correct fermion mass hierarchy we require $\langle \theta \rangle /
M_I \ \sim 0.2$. Thus there is an extra constraint on the models that
\begin{equation}
\frac{M_I}{M_X} = O(1/8).\label{henry}
\end{equation}
We may now search through the models to see if any can make the
gauge couplings unify at $M_X$ {\em and}\/ give an explanatory and
predictive theory of fermion masses. This may be achieved by checking
that Eq.s~\ref{henry},\ref{strings},\ref{Ks} are satisfied and that
the extra matter is in a form to give the correct pattern of light
fermion masses.

\head{Summary}

Intermediate matter is an effective way of obtaining gauge
unification at the string scale. Its mass may come from
hidden sector dynamics or non-renormalisable string-type
operators$^{9)}$.
The intermediate matter can be combined with an abelian family
dependent gauge symmetry to yield the lighter
fermion masses.
It is now possible to construct$^{10)}$ explicit models which
incorporate both a predictive and explanatory theory of fermion masses
and which unify the gauge couplings at the string scale. In this
manner, we may develop a realistic theory
coming from a superstring model.

\end{document}